\documentclass[iop]{emulateapj}
\usepackage{amsmath,amsfonts,amssymb}
\usepackage{footnote}
\usepackage{epsfig}
\usepackage{graphicx,subfigure}
\usepackage{natbib} 
\usepackage{apjfonts}
\usepackage[cm]{fullpage}

\newcommand{\be}{\begin{equation}}

\newcommand{\ee}{\end{equation}}

\newcommand{\bea}{\begin{eqnarray}}

\newcommand{\eea}{\end{eqnarray}}

\newcommand{\beax}{\begin{eqnarray*}}

\newcommand{\eeax}{\end{eqnarray*}}

\newcommand{\ba}{\begin{array}}

\newcommand{\ea}{\end{array}}

\newcommand{\bed}{\begin{description}}

\newcommand{\ed}{\end{description}}

\newcommand{\blc}{\begin{list}{$\circ$}{}}

\newcommand{\blb}{\begin{list}{$\bullet$}{}}

\newcommand{\el}{\end{list}}

\newcommand{\ben}{\begin{enumerate}}

\newcommand{\een}{\end{enumerate}}

\def\lapprox{\mathrel{\hbox{\rlap{\hbox{\lower4pt\hbox{$\sim$}}}\hbox{$<$}}}}

\def\gapprox{\mathrel{\hbox{\rlap{\hbox{\lower4pt\hbox{$\sim$}}}\hbox{$>$}}}}

\begin{document}

\title{Are Decaying Magnetic Fields Above Active Regions\\ Related to Coronal Mass Ejection Onset?}

\author{J. Suzuki}

\affil{Department of Astronomy, University of California, Berkeley, CA  94720-7450}

\author{B.~T. Welsch}

\affil{Space Sciences Laboratory, University of California, Berkeley,  CA 94720-7450}

\author{Y.~Li}

\affil{Space Sciences Laboratory, University of California, Berkeley, CA 94720-7450}

\begin{abstract}

Coronal mass ejections (CMEs) are powered by magnetic energy stored in non-potential (current-carrying) coronal magnetic fields, with the pre-CME field in balance between outward magnetic pressure of the proto-ejecta and inward magnetic tension from overlying fields that confine the proto-ejecta. In studies of global potential (current-free) models of coronal magnetic fields --- Potential Field Source Surface (PFSS) models --- it has been reported that model field strengths above flare sites tend to be weaker in when CMEs occur than when eruptions fail to occur. This suggests that potential field models might be useful to quantify magnetic confinement. One straightforward implication of this idea is that a decrease in model field strength overlying a possible eruption site should correspond to diminished confinement, implying an eruption is more likely. We have searched for such an effect by {\em post facto} investigation of the time evolution of model field strengths above a sample of 10 eruption sites. To check if the strengths of overlying fields were relevant only in relatively slow CMEs, we included both slow and fast CMEs in our sample. In most events we study, we find no statistically significant evolution in either: (i) the rate of magnetic field decay with height; (ii) the strength of overlying magnetic fields near 50 Mm; (iii) or the ratio of fluxes at low and high altitudes (below 1.1$R_{\odot}$, and between 1.1--1.5$R_{\odot}$, respectively). We did observe a tendency for overlying field strengths and overlying flux to increase slightly, and their rates of decay with height to become slightly more gradual, consistent with increased confinement. The fact that CMEs occur regardless of whether the parameters we use to quantify confinement are increasing or decreasing suggests that either: (i) the parameters that we derive from PFSS models do not accurately characterize the actual large-scale field in CME source regions; or (ii) systematic evolution in the large-scale magnetic environment of CME source regions is not, by itself, a necessary condition for CMEs to occur; or both.
\end{abstract}


\section{Introduction}

\label{sec:intro}


In a coronal mass ejection (CME), the Sun launches a hot ($\sim 1$ MK) mass ($\sim 10^{15}$ g) of magnetized plasma at high speed (from a few hundred km sec$^{-1}$ at the slow end to 2000 km sec$^{-1}$ or more at the fast end) into interplanetary space. A solar flare is a transient burst of energy released across a wide range of the electromagnetic spectrum, from radio to X-ray (and in some cases, gamma-ray) wavelengths, over a wide range of energies, from $\sim 10^{31}$ -- $10^{33}$ ergs for X-class flares at the high end to barely detectable X-ray enhancements in weak A-class flares. CMEs are often associated with flares, but the relationship between the two is not one-to-one: some flares produce no CMEs; some (mostly slow) CMEs occur with barely any detectable flare emission; but most fast CMEs are associated with flares, and almost all large (X- and M-class) flares produce CMEs \citep{Andrews2003, Wang2007b}.

CMEs are observed to originate in the low corona. Consideration of available sources of energy there imply that both flares and CMEs must be powered by energy stored in the coronal magnetic field \citep{Forbes2000}: coronal gas densities are so low that thermal energy, gravitational energy, and kinetic energy densities are all too small to supply the energy required to drive CMEs.

The basic physical processes that produce CMEs, however, are not understood. Beyond our scientific interest in CMEs, those directed toward Earth are the primary drivers of severe space weather events. While flares also occasionally produce effects at Earth independent of CMEs --- for instance, ionization of the Earth's upper atmosphere from the enhanced X-ray flux can interfere with radio communications --- the biggest terrestrial effects arise from CMEs \citep{Gosling1993}. Consequently, the practical capability to predict CMEs would be quite useful. Because the physical processes at work in CMEs are not understood, however, we are still at the stage of seeking empirical correlations between solar observations and the occurrence of CMEs. 

The CME process can be understood in terms of the interplay between magnetic pressure and tension forces in the corona. (The same reasoning used to argue that CMEs must be powered by magnetic energy also implies that, in the corona, non-magnetic forces --- such as gas pressure gradients and gravity --- are not the primary drivers of CME dynamics.)  The pre-CME field is thought to be in balance between outward magnetic pressure and inward confinement by magnetic tension.

Essentially, CME initiation requires an increase in outward magnetic pressure, or a decrease in confinement, or both. In the tether-cutting model of \citet{Moore2001}, for instance, field lines that arch over the proto-ejecta can reconnect underneath it, converting downward tension into upward tension, which both increases the outward force and decreases the inward force; therefore the system will expand outward. As more underlying field lines reconnect, the outward forces increase. If the increase in outward forces is insufficient to overcome the confinement by magnetic tension, then no eruption occurs. If, however, the outward forces overcome the restraining tension, then an eruption occurs. In the breakout model of \citet{Antiochos1999a}, ``breakout reconnection'' above the proto-ejecta decreases overlying magnetic tension, causing the proto-ejecta to rise, which initiates ``flare reconnection'' underneath it, leading to CME initiation.

In contrast to these non-ideal models of CME initiation, ideal MHD instabilities might also produce CMEs. \cite{Kliem2006} suggested the torus instability could explain the CME process: if the outward force from magnetic pressure in the proto-ejecta decays with radial distance from the photosphere more gradually than the inward magnetic tension of the overlying field that confines the proto-ejecta, then any outward perturbation of the proto-ejecta will be unstable, and an ejection will occur. In a given configuration, variations in either the field of the proto-ejecta or the confining field (or both) could enable the torus instability to occur.

One key property of flares and CMEs is that they occur along polarity inversion lines (PILs) of the photospheric magnetic field (e.g., \citealt{Falconer2006}). Observers report that the orientations of H-$\alpha$, EUV (extreme ultraviolet), and SXR (soft X-ray) structures low in the solar atmosphere above PILs typically deviate from directions expected from potential field models, implying electric currents are present; but higher in the corona, EUV and SXR loops typically appear nearly potential \citep{Martin1996}.

These observations of potential-like field configurations far above PILs suggest potential field models could be studied to understand the role of magnetic field confinement in CMEs. \citet{Wang2007b} studied potential field models associated with eight X-class flares, four with and four without CMEs, computing the ratio of flux at low altitude in the corona, below $1.1 R_\odot$, to the flux from $1.1 R_\odot - 1.5 R_\odot$ for each case. They found flares with {\em less} confinement relative to magnetic pressure (a {\em higher} ratio of low flux to high flux) were more likely to erupt. In addition, \citet{Liu2008} analyzed several failed eruptions and CMEs, fitting the decrease with radius of horizontal field strength in global potential models, and found that field strengths decreased faster with radius in successful eruptions. He also found average field strengths at a height of $42$ Mm were higher in cases with failed eruptions.

Here, we also investigate the magnetic structure above eruptive PILs in potential magnetic field models, to look for patterns in the evolution of the large-scale model fields around the times of CMEs.
In the next section, we discuss how we chose our events, then we discuss our methods to analyze the data in \S \ref{sec:methods}, look for evidence of systematic magnetic field evolution within the data in \S \ref{sec:results}, and finally discuss our results in \S \ref{sec:conclusion}.


\section{Data}
\label{sec:data}

For this research, we selected 10 CMEs over a 13 year span, from 1997 to 2010. Our goal was to compile an event sample commensurate in length with those used by \citet{Liu2008} and \citet{Wang2007b}. To select the events we analyzed, we first manually identified source active regions (ARs) for 100 CMEs that were within a box measuring 60$^{\circ}$ by 60$^{\circ}$ at disk center. CME dates and times were taken from SOHO/LASCO CME catalog\footnote{http://cdaw.gsfc.nasa.gov/CME\_list/UNIVERSAL/text\_ver/univ\_all.txt}. The RHESSI \footnote{http://hesperia.gsfc.nasa.gov/hessidata/dbase/hessi\_flare\_list.txt} and NGDC X-ray flare lists\footnote{ftp://ftp.ngdc.noaa.gov/STP/space-weather/solar-data/solar-features/solar-flares/x-rays/goes/}, combined with University of Hawai'i's Institute for Astronomy active region maps\footnote{http://kopiko.ifa.hawaii.edu/ARMaps/Archive/} (compiled from Solar Region Summaries released by NOAA's Space Environment Center), were used to identify CME source active regions. The requirement that CME source regions be unambigously identifiable in data from the CME and flare catalogs was a restrictive criterion. Starting from the CDAW CME catalog, we recorded each usable CME event until we found 100. This compilation was not meant to be exhaustive; we sought, instead, only a sample of CMEs for which source active regions had been identified.

We then distilled this list down to 10 by selecting the most isolated, compact, and near-disk-center events. 
We chose ARs which were spatially isolated as well as compact; these constraints reduce the likelihood that our magnetic field models will be strongly affected by neighboring active regions. We also chose a sample that includes both fast and slow CMEs to see if the magnetic environment of CME source regions plays a strong role in either fast or slow populations. We defined a fast CME to be above 800 km sec$^{-1}$; because most CMEs are slower than this, we have more slow CMEs in our sample. The 10 ARs which best fit these criteria were used for modeling in this research.
Basic properties of our events are presented in Table \ref{tab:artable}.
Our events consist of generally weaker flares than those of Wang \& Zhang (all X-class) and Liu (several X- and M- classes, in addition to a few C-class flares). We note that the May 1997 event has been studied previously by several researchers, e.g. \citet{Li2010} and \citet{Liu2008}. 

\begin{deluxetable*}{lcrrrrrr}
\tabletypesize{\scriptsize}
\setlength{\tabcolsep}{1.0in} 
\tablecaption{CME and Active Region Table}
\tablewidth{\textwidth}
\tablehead{
  \colhead{NOAA}
& \colhead{Lat./Long.}
& \colhead{CME Date}
& \colhead{Flare Time,}
& \colhead{CME Time}
& \colhead{CME Speed}  
& \colhead{GOES Flare} 
& \colhead{$B_t$ at $t=0$\tablenotemark{b}} 
\\ 
  \colhead{AR No.} 
& \colhead{(deg)} 
& \colhead{yyyy/mm/dd}
& \colhead{$t=0$ (UT)}
& \colhead{(UT)}
& \colhead{(km s$^{-1}$)}  
& \colhead{Class} 
& \colhead{(G)} 
}
\startdata
8038
& N21W09
& 1997/05/12
& 05:26
& 05:30
& 464
& C1.3
& 9\\
10311  
& S12E04
& 2003/03/13
& 02:13
& 02:54
& 1021
& C1.3
& 10\\
10696
& N09E05
& 2004/11/06
& 00:57
& 01:31
& 818
& M5.9
& 30\\
10715
& N05E06
& 2005/01/03
& 04:22
& 05:06
& 363
& C3.8
& 7\\
10775 
& N06W07
& 2005/06/12
& 02:36
& 02:36
& 590
& C3.5
& 2\\
10822 
& S07E02
& 2005/11/19
& 23:38
& 00:30
& 231
& C4.1
& 35\\
10956
& N03E00
& 2007/05/19\tablenotemark{a}
& 13:02
& 13:24
& 958
& B9.5
& 10\\
10960
& S07W02
& 2007/06/07
& 17:20
& 17:30
& 263
& C1.1
& 40\\
10977
& S05W07
& 2007/12/07
& 04:41
& 05:30
& 209
& B1.4
& 2\\
10978 
& S09W10
& 2007/12/12
& 05:32
& 06:30
& 305
& B2.4
& 40\\
\enddata 
\tablenotetext{a}{While the CME appeared in LASCO on November 19, the flare occurred the previous day.}
\tablenotetext{b}{The mean $B_t$ at $r = 46.5$ Mm, averaged over the photospheric active-region mask; see \ref{sec:btransverse}.}
\label{tab:artable}
\normalsize
\end{deluxetable*}


\section{Methods}
\label{sec:methods}

We build upon previous work by looking for any systematic time evolution in parameters like those studied by \citet{Liu2008} and \citet{Wang2007b} prior to CMEs. These authors compared CMEs and failed eruptions in different active regions, and found that in potential field models of active regions that produced CMEs (cf., failed eruptions), either overlying field strengths decreased more rapidly with height or there was less total flux overlying the PIL. These authors suggest both that the confinement by magnetic fields above active regions is a key factor in determining whether an eruption can occur, and that potential field models can usefully quantify this confinement. Neither paper discussed any prediction about relationships between time evolution in overlying fields and the occurrence of a CME within an active region. 
We conjecture, however, that if overlying fields do in fact play a
major role in confining pre-CME fields, then it is reasonable to
expect overlying fields to weaken around the times of CMEs. The
occurrence of CMEs despite lack of weakening in overlying fields in
potential models would suggest that either (i) confinement by
overlying fields is not a key factor in CMEs; or (ii) potential field
models do not accurately quantify confinement; or both.
\citet{Liu2008} also mentions that the behavior of transverse magnetic field strength at low heights may also play a role in CME onset. We test our conjecture in our sample of CMEs, to see if these parameters of the CME magnetic environment change systematically from about 48 hours before to 24 hours after each CME. 

We investigate model fields up to 24 hours after the CME event for two reasons. First, we wanted to know if the evolution of the model field around the time of the CME was part of a longer-term trend. If a CME occurs in the middle of an episode of persistent flux emergence, but there is no particular temporal structure in the evolution of the large-scale model field around the time of the CME, then we have learned something: while flux emergence might destabilize the coronal field to produce CMEs, this property of the model potential field cannot be used to detect such destabilization. Second, the additional post-CME model coronal fields are based upon independent measurements, and can therefore reveal whether changes in model field parameters near the CME are temporary fluctuations, and might therefore be ascribed to noise.

To model the magnetic environment of CMEs, we use global Potential Field Source Surface (PFSS) models produced with the codes developed by \citet{Schrivjer2003} from magnetograms observed by \textit{SOHO}/MDI\footnote{http://soi.stanford.edu/magnetic/index5.html} \citep{Scherrer1995}. This model's photospheric fields are hybrids of recent and past observations, combined with a flux transport model. Recent 96-minute-cadence magnetograms are used for the visible disk, and the flux transport model includes diffusion and a statistical treatment of small-scale flux emergence for the far-side boundary condition. 
Except where explicitly stated otherwise, we use PFSS models computed
with spherical harmonic coefficients up to $\ell_{\rm max} = 192$.
The models were computed 6 hours apart, maximizing time resolution while minimizing information redundancy. The photospheric Alfv\'en speed --- of order 1 km s$^{-1}$ --- governs the timescale of photospheric magnetic evolution. In contrast to the spatial resolution of the magnetograms assimilated into the models (about two 1.4 Mm pixels at disk center), the spatial resolution of these PFSS models is about two heliocentric degrees, or $\sim$ 24 Mm near disk center, implying a photospheric Alfv\'en crossing time for model pixels of about 6 hours. (In the corona, the Alfv\'en speed is of order 1000 km s$^{-1}$, implying dramatically faster evolutionary timescales.)  Since the potential field models were only constructed every 6 hours, we matched the CME event times to the time of the closest model field, which we define to be $t=t_0 = 0$.

Since we are interested in how our results compare to \citet{Liu2008}, we followed his approach and fitted transverse field strengths over a range of radii similar to his, at heights of \{46.5, 59.2, 72.3, 85.8, 99.9, 114.5\} Mm, assuming the transverse field is described by a power law in this range. Transverse field strength is defined as:

\be
B_t = \sqrt{B_{\phi}^{2} + B_{\theta}^{2}}~,
\ee
where $B_\phi$ and $B_\theta$ are the orthogonal components of the magnetic field parallel to the Sun's surface. 

We are interested in the time evolution of the magnetic field decay rate, $\gamma$, as a function of height, henceforth assumed to follow a power law in radius with exponent $\gamma$, where
\be
B_t \propto r^\gamma ~,
\ee
and $r$ is the distance above the solar surface. A more negative value corresponds to a magnetic field which is decreasing in strength more steeply with radius. In order to compute the best estimate of $\gamma$, we tried two approaches: averaging the magnetic field at each height within an AR then fitting a $\gamma$ to the averages (Method 1); and calculating a $\gamma$ for each pixel within an AR and then averaging the $\gamma$'s (Method 2). Furthermore, we apply Methods 1 and 2 for two separate areas: the entire AR itself, as well as just the PIL. We analyzed the model fields near PILs separately because PILs are source regions for CMEs.

We use a linear, least-squares fit \citep{Press1992} to the log-log values of magnetic field strength over radial height to find power-law exponents. When fitting the logarithms, we set the origin of the independent variable, $\log( r / r_0)$, to be between the third and fourth radial values. This enables us to plot lines depicting the fitted $\gamma$'s with errors more clearly without reference to variations in the fitted $y$-intercept, which we did not investigate.

For Method 1, uncertainties on $\gamma$ are derived from the standard error in the estimate of the average magnetic field at each height through the fitting procedure. To estimate the uncertainty in the average transverse field at a given radius $r_i$, we first compute the standard deviation $\sigma_{Bt}$ of $B_t$ at $r_i$ over pixels corresponding to the photospheric mask (AR or PIL), and then divide $\sigma_{Bt,i}$ by $\sqrt{N_{\rm pix}}$, where $\sqrt{N_{\rm pix}}$ is the number of pixels in the mask. This treats $B_t$ from each pixel in the mask as an independent estimate of the mean field strength at $r_i$. An array of these uncertainties, one per each fitted radial point, is passed to the least-squares fitting
procedure, where it is used in weighting the points to be fit. The procedure computes the square of the uncertainty in fitted slope $\gamma$ as one over the variance of $\{r_i/(\sigma_{Bt,i}/\sqrt{N_{\rm pix}})\}$, multiplied by the sum over $i$ of $\{ 1/(\sigma_{Bt,i}/\sqrt{N_{\rm pix}}) \}$. For Method 2, the uncertainty in $\gamma$ at a given time is the standard error in the estimate using the fits to $\gamma$ from each pixel: the standard deviation of the set $\{\gamma_j \}$, divided by $\sqrt{N_{\rm pix}}$, where $\gamma_j$ is the linear fit to $B_t(r)$ in the $j$-th pixel, and $N_{\rm pix}$ is the number of pixels in the appropriate mask.

\begin{figure}[!htb] %
\vspace{-75pt}
\centerline{%
\includegraphics[width=4.0in]{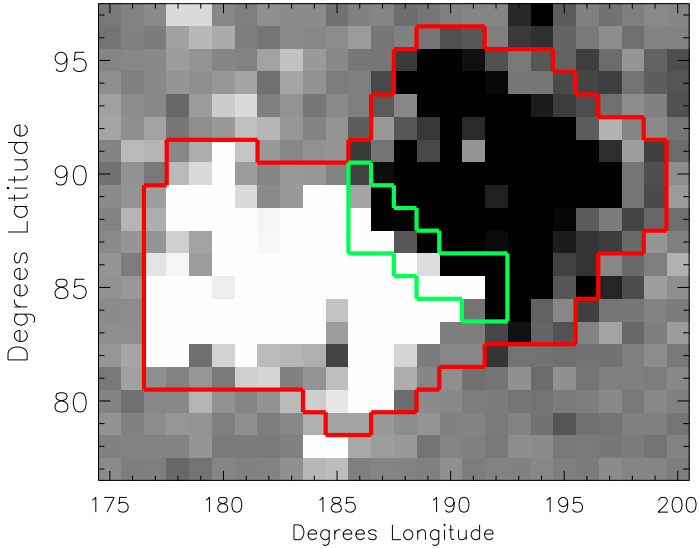}}
\caption{Pixels used in the June 07, 2007 AR. The red and green outlined regions represent the AR and PIL mask, respectively. The background grayscale shows the radial magnetic field (white positive, black negative) at the photosphere in the PFSS model. The AR mask used in finding $\gamma$ is defined to be the pixels whose absolute radial field strength is above 25 Mx cm$^{-2}$. The PIL mask is computed by dilating the masks of positive and negative polarities with absolute radial field strengths greater than 50 Mx cm$^{-2}$ by 1 pixel and finding where they overlap. The grayscale saturation level of the image is set to $\pm$ 50 Mx cm$^{-2}$.}
\label{fig:soloar}
\end{figure}

\begin{figure*}[!htb]
\centerline{%
\includegraphics[width=0.84\textwidth]{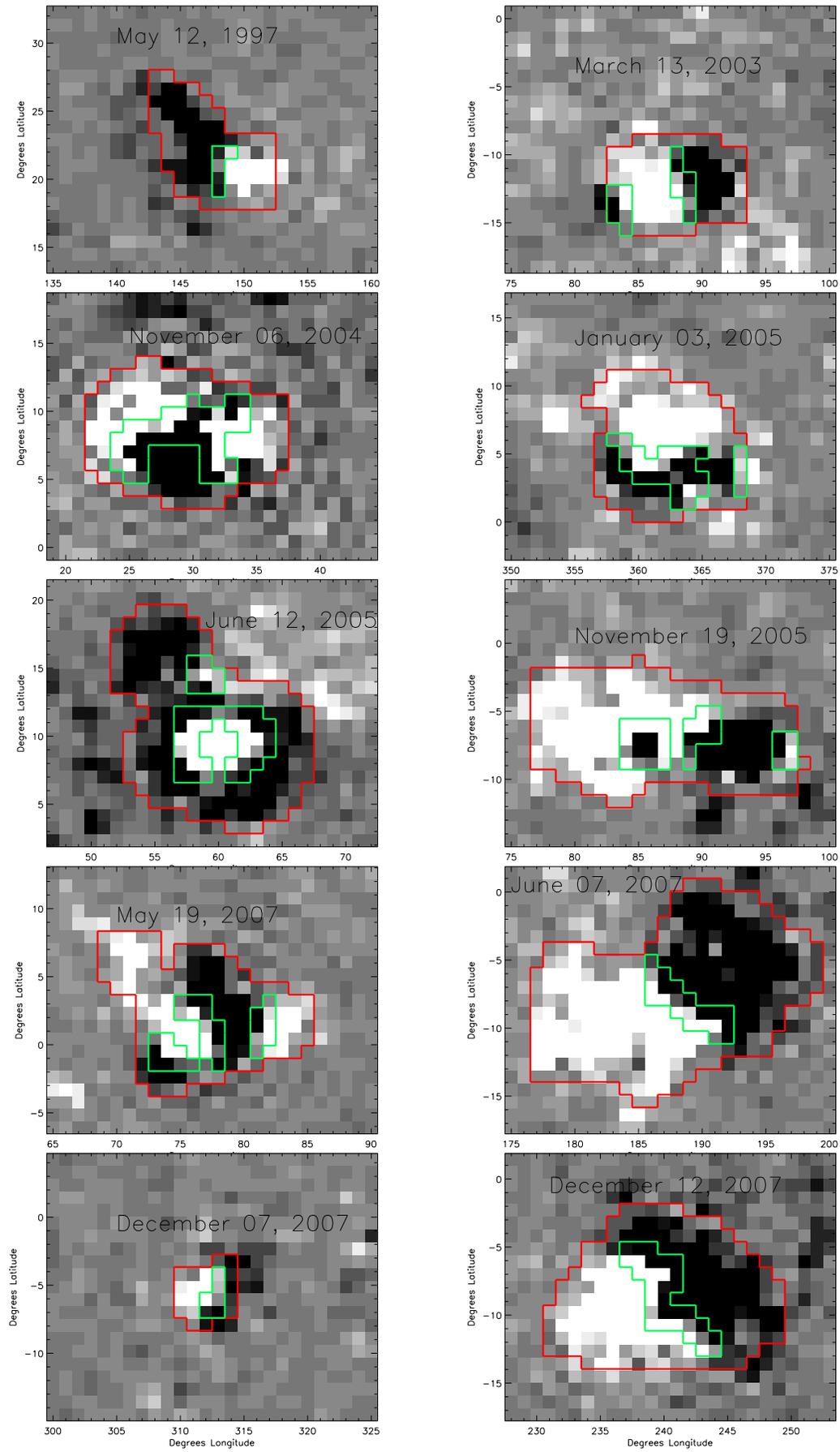}}
\label{tab:newmagnetomaps}%
\vspace{10pt}
\caption{Magnetograms of source ARs. The red and green outlines correspond to pixels used for the entire AR and the PIL-region, respectively. Grayscale saturation levels are $\pm$ 50 Mx cm$^{-2}$.}
\end{figure*}

Figure \ref{fig:soloar} shows how we mask the AR and select pixels from which to compute $\gamma$. The background grayscale shows the radial magnetic field from the PFSS model at $r=0$, with white representing positive fields and black representing negative fields. We computed the AR mask by identifying pixels with an absolute radial magnetic field strength above 25 Mx cm$^{-2}$. We identified PIL masks following the method of \citet{Schrijver2007} and \citet{Welsch2008b}: masks of pixels from each polarity stronger than a threshold are dilated by a pixel and multiplied together; regions of overlapping pixels are thereby identified as the PIL mask. A higher threshold, 50 Mx cm$^{-2}$, is used in finding the PIL mask; starting with too low a threshold leaves too many pixels in strong field regions of either polarity. Similar magnetograms of all ARs in our sample are shown in Figure 2.

\begin{figure*}[!htb]%
\centerline{%
\includegraphics[width=.8\textwidth]{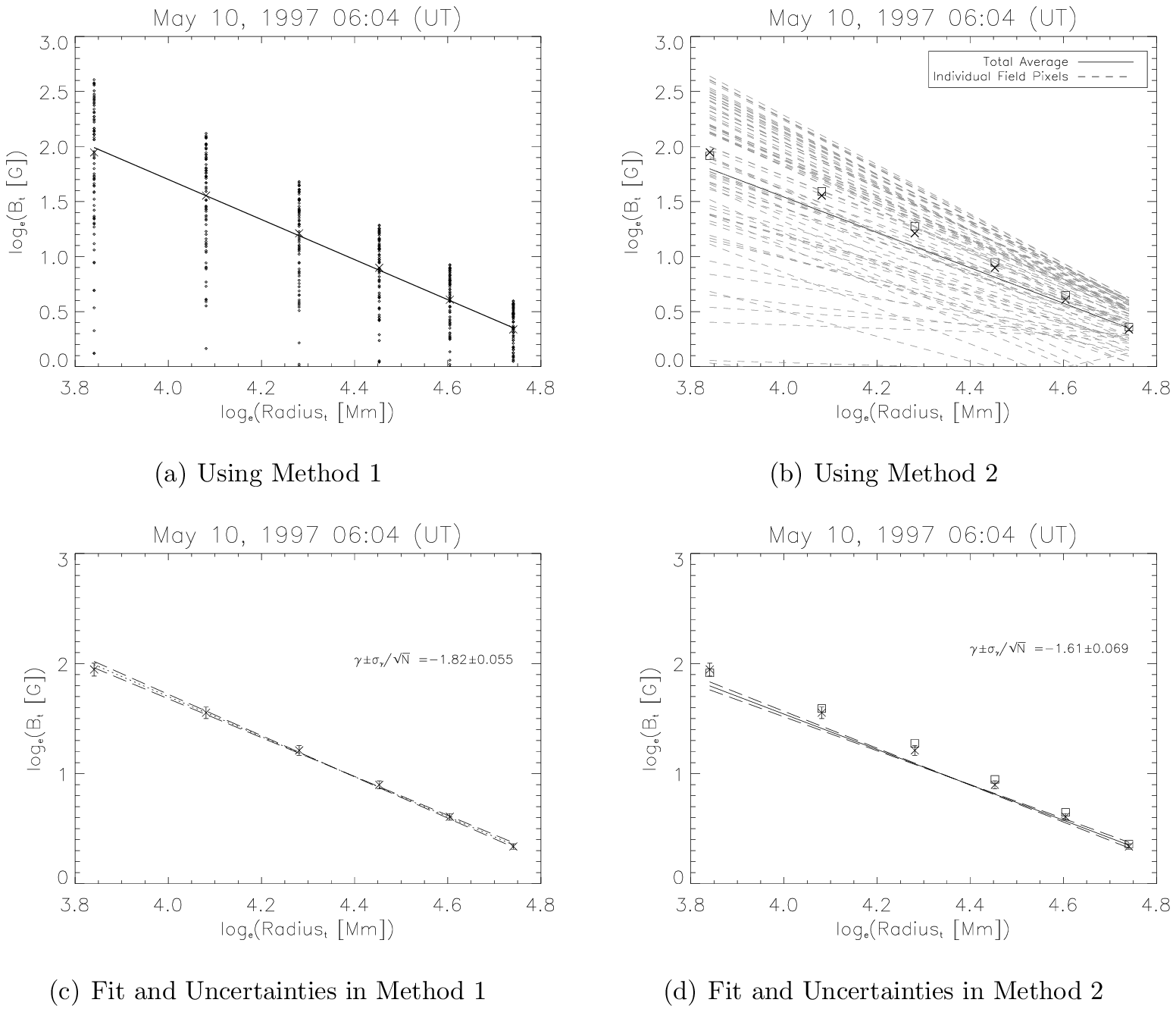}}%
\caption[]{Comparison of different Methods 1 and 2 used to calculate $\gamma$. In (a), the black dots represent transverse magnetic field strengths at each radius, and the average values ($\times$) are fitted. In (b), the dashed lines represent the fitted power laws for field strength above each pixel, and the solid line is the average of the fitted slopes. We look at values of the fitted power-law exponents (the fitted slopes) of the solid lines over time for each event in our sample. In (c) and (d), we plot fitted slopes varied by the estimated uncertainties. In all four panels, the $\times$ denote the average field. The boxes in (b) and (d) represent the median of the magnetic field values at each radius. While the distribution of field values is wide, the median is close to the average.}%
\label{fig:all}%
\end{figure*}

\begin{figure*}[!htb]%
\centerline{%
\includegraphics[width=.95\textwidth]{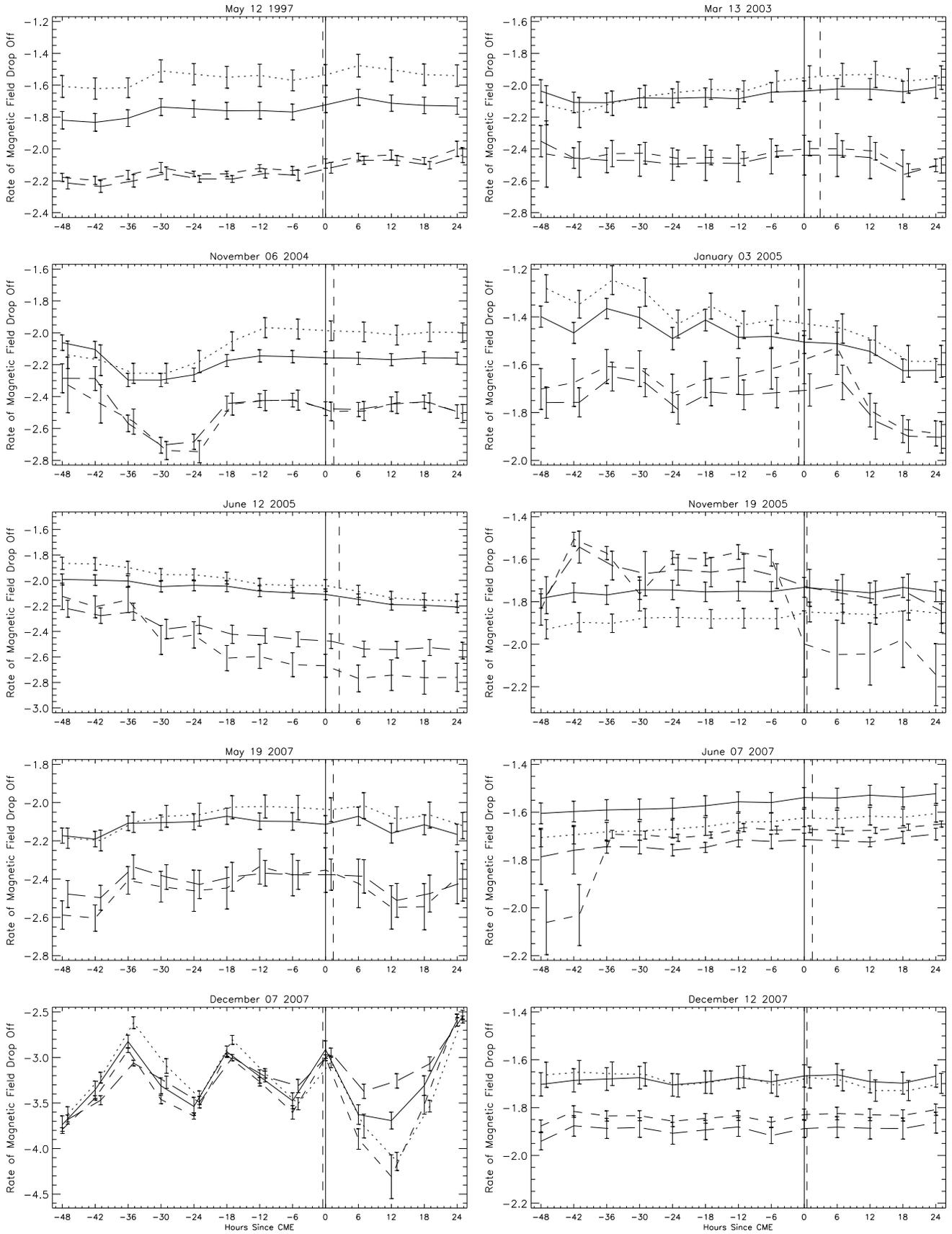}}%
\caption{Time evolution of $\gamma$'s. The solid line is Method 2 (AR), the dotted line is Method 1 (AR), the long dashed line is Method 2 (PIL) and the short dashed line is Method 1 (PIL). For most events, Methods 1 and 2 give similar answers for AR, and for PIL --- but PIL $\gamma$'s are significantly more negative than AR $\gamma$'s. A more negative $\gamma$ corresponds to more rapid decay of the magnetic field with height. Some error bars and data points are slightly offset in time to avoid overlap. The solid vertical line corresponds to $t=0$ and the dashed vertical line corresponds to the CME time.}
\label{fig:finalplotsv2}
\end{figure*}

When choosing to analyze the entire AR, the spread of magnetic field strengths is typically very wide (see panel (a) of Figure \ref{fig:all}). Method 2 nearly always results in a $\gamma$ with a smaller magnitude than Method 1. In addition to this, it also results in larger uncertainties in $\gamma$. While Method 1 fits closer to the average magnetic field, this does not necessarily mean it is a better estimate of the decay rate in an AR's overlying magnetic field. Method 2 takes into account field data which stray from the average. While the individual points have an insignificant impact on the final result of Method 1, they appear in Method 2 by shifting the final $\gamma$ down. Each method offers a slightly different perspective on field structure. We applied two methods of calculating $\gamma$'s in an attempt to test the robustness of our computation. \textit{As shown in Figure \ref{fig:finalplotsv2}, we found that the main factor which determined the value of $\gamma$ was not the computation method, but instead which part of the AR we looked at.} 

As a caveat, due to lack of resolution in the magnetic models, for a few events there are very few PIL pixels; in some, there are less than 10 pixels in the PIL masks.

\begin{figure}[!h]%
\vspace{5pt}
\centerline{%
\includegraphics[width=4.5in]{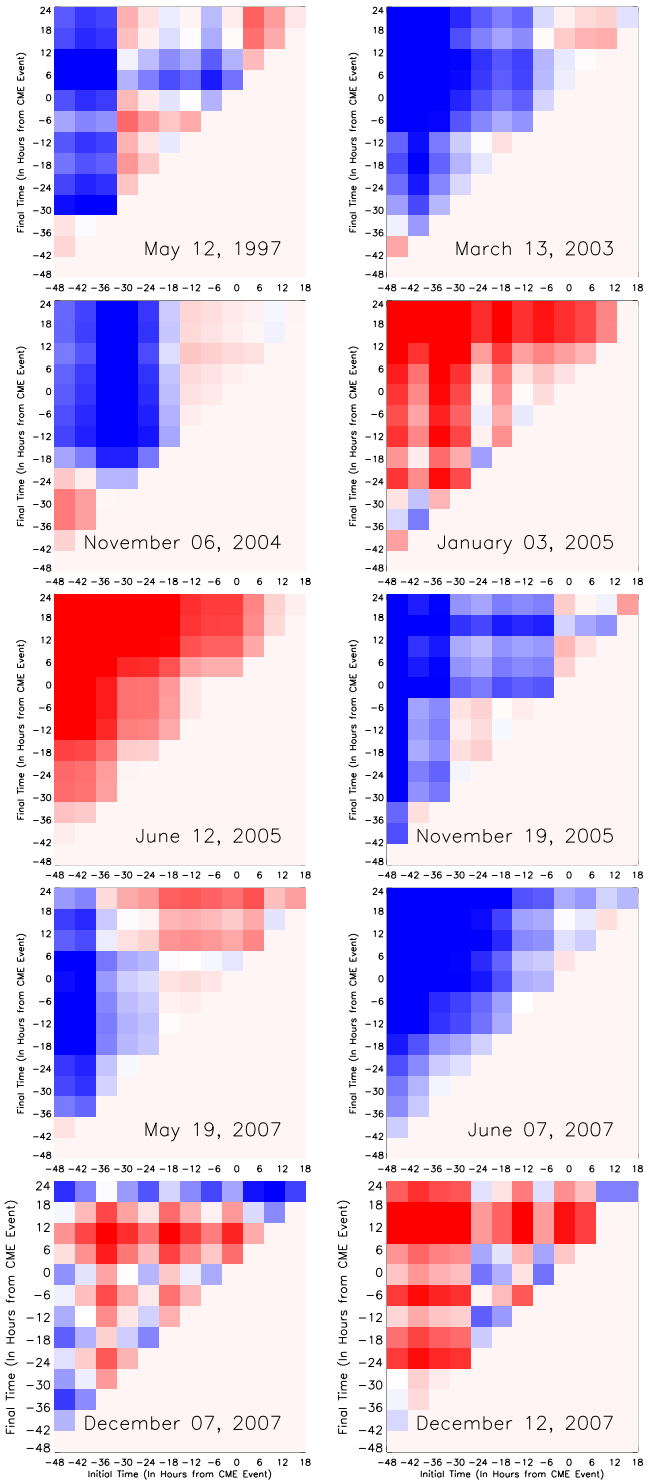}}%
\vspace{10pt}
\caption{$\Delta \gamma$ between all possible pairs of $\gamma_{\textrm{AR}}$'s. The saturation level is $\pm 2$, scaled by the standard deviation in $\gamma_{\textrm{AR}}$ for each event, with negative changes represented as red and positive changes as blue. Blues predominate, implying that, over time, field decay rates typically grow more positive, i.e., fields decay {\em less} rapidly with height.}
\label{tab:trianglegamma}
\end{figure}

\begin{figure*}[!h]%
\centerline{%
\includegraphics[width=.95\textwidth]{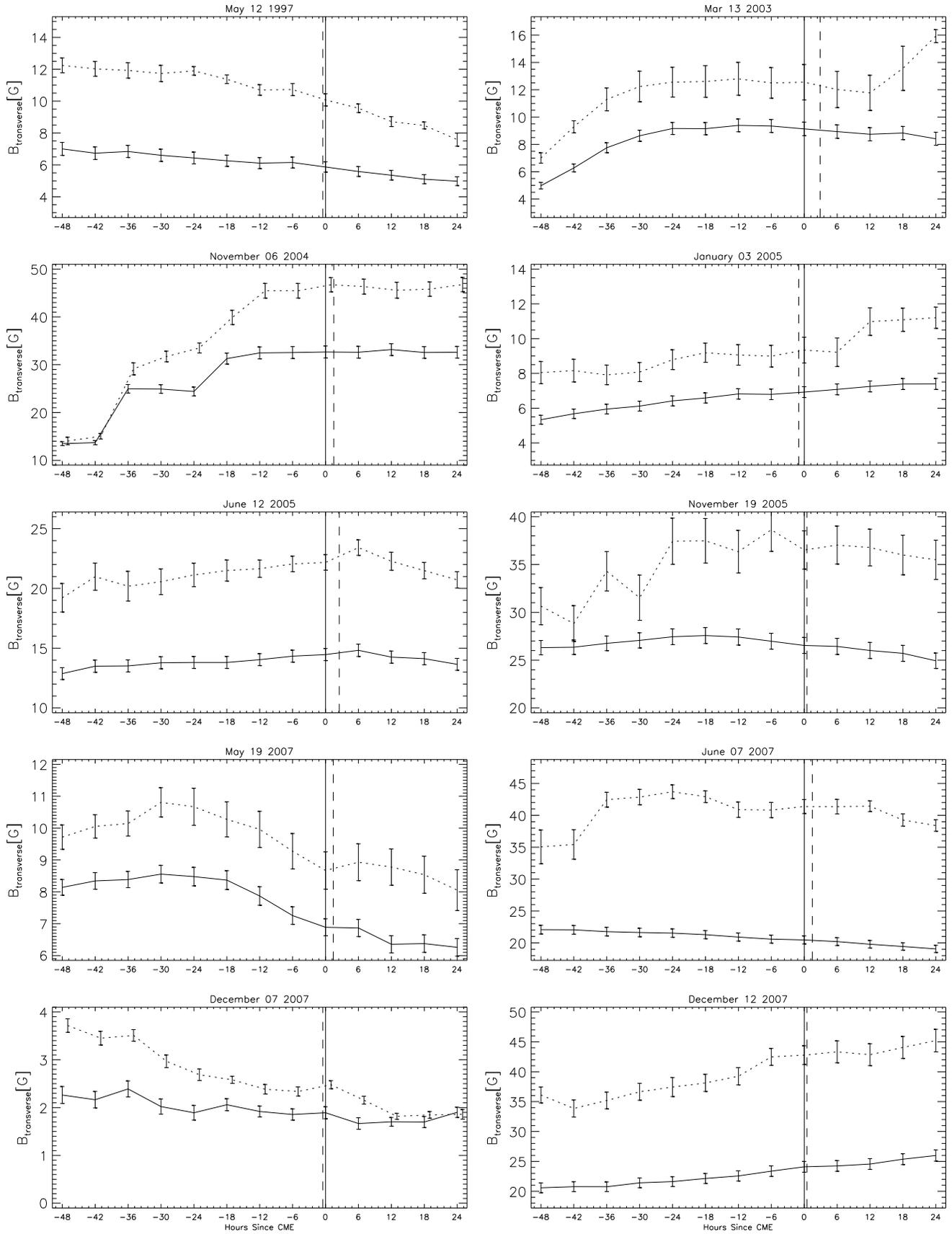}}%
\caption{Time evolution of the averaged transverse magnetic field $B_t$ at 46.5 Mm. The solid and dotted lines correspond to fields computed from ARs and PILs respectively. Some error bars and data points are slightly offset in time to avoid overlap. The solid vertical line corresponds to $t=0$ and the dashed vertical line corresponds to the CME time. }
\label{fig:bevolution}
\end{figure*}

\begin{figure}[!h]%
\vspace{5pt}
\centerline{%
\includegraphics[width=4.5in]{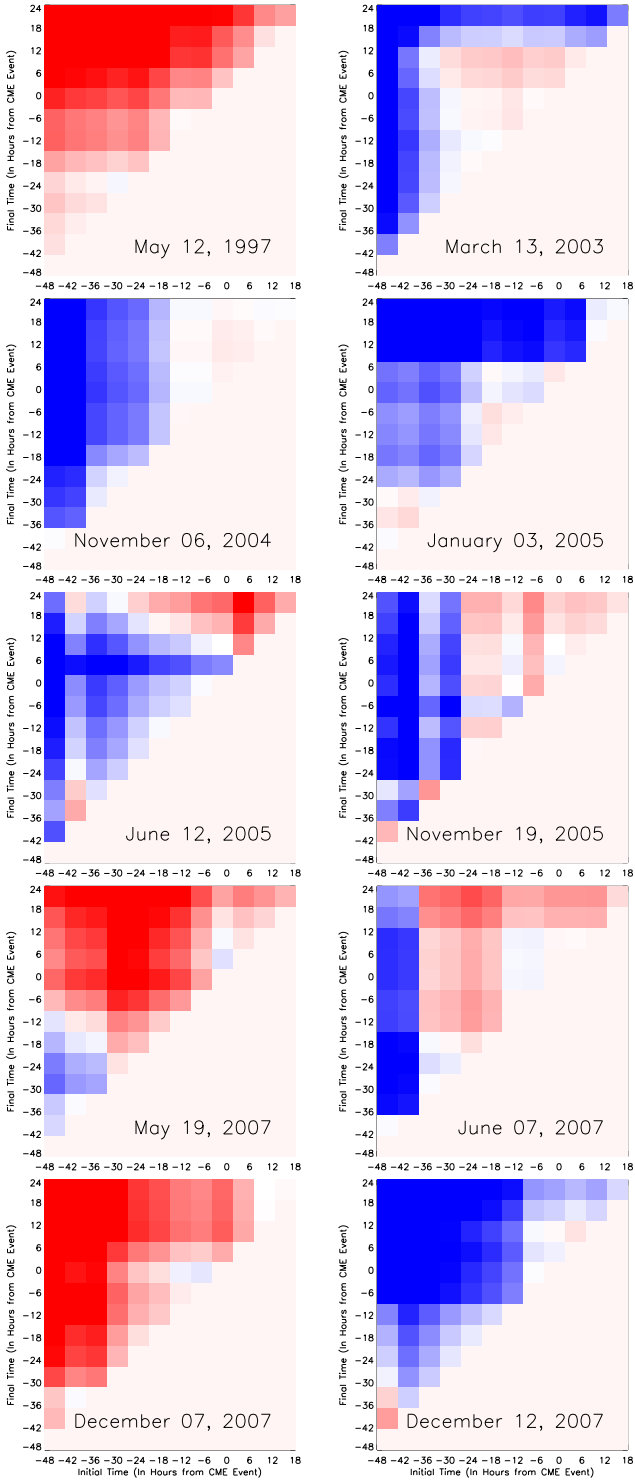}}%
\vspace{10pt}
\caption{$\Delta B_{t\textrm{PIL}}$ between all possible pairs of $B_{t\textrm{PIL}}$'s. The saturation level is $\pm 2$, scaled by the standard deviation in $B_{t\textrm{PIL}}$ for each event, with negative changes represented as red and positive changes as blue. Blues predominate, implying $B_{t\textrm{PIL}}$ tends to increase in time.}
\label{fig:trianglebarraypil} 
\end{figure}

\begin{figure*}[!h]%
\centerline{%
\includegraphics[width=.95\textwidth]{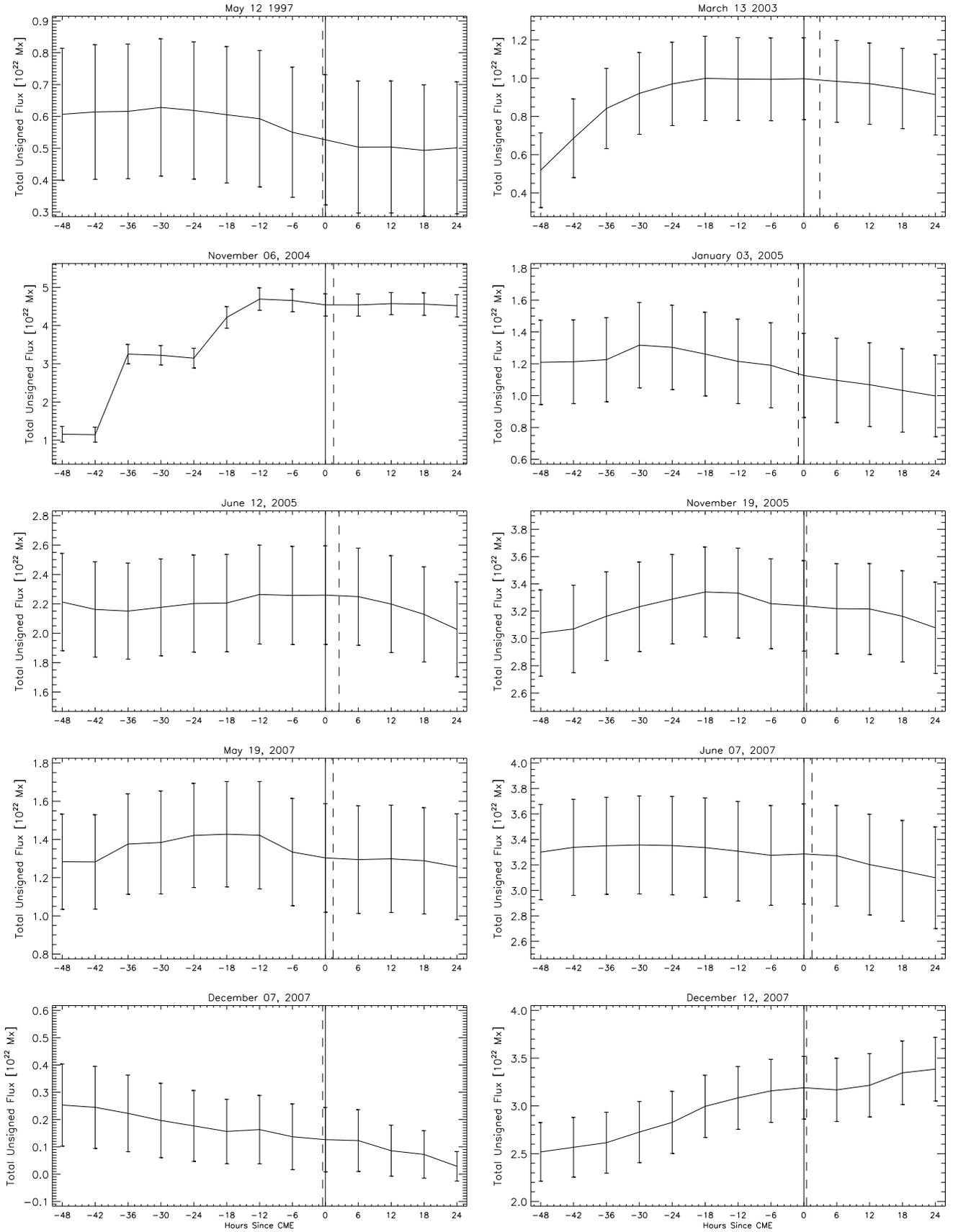}}
\caption{Time evolution of the total unsigned flux in each event, summed over each AR. Error bars were derived assuming uncertainties of 20 Mx cm$^{-2}$ in the MDI magnetograms used to produce the models. The solid vertical line corresponds to $t=0$ and the dashed vertical line corresponds to the CME time.}
\label{fig:unsignedflux}
\end{figure*}

\section{Results}
\label{sec:results}


\subsection{Evolution in $\gamma$}
\label{sec:gamma}

\begin{deluxetable}{lrrrrr}
\tabletypesize{\scriptsize}
\tablecaption{Evolution of $\gamma$}
\tablewidth{\columnwidth}
\setlength{\tabcolsep}{0.2in} 
\tablehead{
  \colhead{\hspace{-.07in}Initial Time\tablenotemark{a}}
& \colhead{\hspace{-.1in}Final Time\tablenotemark{a}}
& \colhead{\hspace{-.1in}$\textstyle \Delta \gamma_{\textrm{AR}} > 0$\tablenotemark{b}}
& \colhead{\hspace{-.1in}$\Delta \gamma_{\textrm{AR}} < 0$}
& \colhead{\hspace{-.1in}$\Delta \gamma_{\textrm{PIL}} > 0$}
& \colhead{\hspace{-.1in}$\Delta \gamma_{\textrm{PIL}} < 0$}  
}
\startdata
-6
& 0
& 6 (1)
& 4
& 7 (1,1)
& 3 (1)\\
-12  
& 0
& 6
& 4 (1)
& 5 (1)
& 5 (1)\\
-24
& 0
& 7 (1)
& 3
& 8 (2,1)
& 2 (1)\\
-48
& 0
& 7 (1,1)
& 3 
& 7 (2,1)
& 3 (1,1)\\
-12
& 12
& 4 
& 6 (1,1)
& 3
& 7 (2,1)\\
-24
& 24
& 6 (1,1)
& 4
& 6 (2,1)
& 4 (2,1)\\
-48
& 24
& 7 (1,1)
& 3 (2)
& 5 (2,1)
& 5 (2,1)\\
\enddata 
\tablenotetext{a}{Measured in hours since closest magnetogram to CME event}
\tablenotetext{b}{Most values are not significant at a 1$\sigma$ confidence level; the number of events for which values are significant are listed in parentheses with the first and second entries corresponding to 1$\sigma$ and 2$\sigma$ significance, respectively.}
\label{tab:gammatable}
\normalsize
\end{deluxetable}

The purpose of this research is to see if there is any relation between evolution of $\gamma$ and a CME event. Figure \ref{fig:finalplotsv2} reveals no strong pattern in $\gamma$'s: in most cases, $\gamma$ remains relatively unchanged, with no strong tendencies (relative to error bars) for increases or decreases. We note that our values for $\gamma$ are similar to those determined by \citet{Liu2008}, who found $\gamma$'s in the range of 1.51 -- 2.25 over a range of radii (42 -- 105 Mm) similar to ours. (In his Fig. 3, \citealt{Liu2008} shows a faster fall-off of field strength with radius beyond about 400 Mm.) For June 07 2007, we saw a slight increase in $\gamma$ (implying a more gradual decay of transverse field with radial height), while for June 12, 2005 we saw a slightly stronger decrease in $\gamma$ (steeper decay of field strength with radius). (Alone among our events, the December 07 event  exhibits strong back-and-forth fluctuations.) A key point to keep in mind is that CMEs can evidently occur for any type of evolution in $\gamma$. Furthermore, we found no significant correlation between $\gamma (t_0)$ and either CME speed or flare size. For the entire set of events considered as a whole, we do not see evidence of systematic decreases in $\gamma$'s around CME onset, regardless of whether $\gamma$ is computed with Method 1 or 2, or for whole ARs or only PILs. 

We note that the $\gamma$'s computed from PILs are consistently more negative than their AR-computed counterparts. One possible explanation for this is due to the nature of PILs themselves. The PIL lies in the region of the photosphere between the two polarities, meaning regions of corona above PILs are farthest away from photospheric source regions, resulting in weak potential fields (which scale as the square of the inverse distance). However, this behavior only affects the value of $\gamma$, and not its evolution over time. 

To investigate whether or not there exists a more subtle pattern in the evolution of $\gamma$, in Table \ref{tab:gammatable} we show changes in $\gamma$ determined using Method 1, for several time intervals near each CME event. We chose four time intervals prior to CMEs to look for systematic, pre-CME changes in $\gamma$. \citet{Longcope2005} and \citet{Schrijver2005} both found characteristic time scales of coronal energy storage of about 24 hours, suggesting this might be a characteristic time relevant to CMEs. Our pre-CME intervals were chosen to sample near this characteristic time scale. In addition, we also computed field changes over one, two, and three day intervals that straddled each CME. We find that most changes in $\gamma$ are not statistically significant at the 1-$\sigma$ level; changes that are significant at the 1- and 2-$\sigma$ level are shown in parentheses. We only show values of $\gamma$ from Method 1 in Table \ref{tab:gammatable} because both methods had similar results for most time intervals.

Table \ref{tab:gammatable} reveals that, while most of the changes in $\gamma$ are not statistically significant, for $\Delta \gamma_\textrm{AR}$, there is a preponderance of positive changes in $\gamma_\textrm{AR}$: for six out of seven intervals, positive $\Delta \gamma_\textrm{AR}$'s outnumber negative $\Delta \gamma_\textrm{AR}$'s. To test whether this is significant or not, we compared our result with the null hypothesis: that changes in $\gamma_\textrm{AR}$ are random. Since the evolution of $\gamma_\textrm{AR}$ is observed to be either positive or negative, the signs of $\gamma_\textrm{AR}$'s under the null hypothesis can be modeled with the binomial distribution. We first calculated the probability that a positive bias in $\Delta \gamma_\textrm{AR}$ (more than five events with a positive change) would occur in one time interval (0.37); we then calculated the likelihood that the same bias would occur in at least six out of seven intervals (0.01). Consequently, we argue that our results are inconsistent with random changes in $\gamma_\textrm{AR}$ (which corresponds to the signs of $\Delta \gamma$ also being random): $\gamma$ tends to become more positive, implying {\em less rapid} decay of $B_t$ with height as time goes on. There is also a predominance of positive $\Delta \gamma_{PIL}$'s in Table \ref{tab:gammatable}, but not as great as that among the $\Delta \gamma_\textrm{AR}$'s. 

We note that, because the same magnetic model appears in multiple time intervals, a few random excursions in $\gamma$ at the times that we selected to difference $\gamma_\textrm{AR}$ could have introduced a spurious pattern into our results. We therefore evaluated $\Delta \gamma_\textrm{AR}$ for all possible time intervals in our dataset. In Figure \ref{tab:trianglegamma}, we show positive changes in blue and negative changes in red, for each event. The saturation in the color scale is set to $\pm$ 2, where $\Delta \gamma_\textrm{AR}$ for each event is normalized by the standard deviation in $\gamma_\textrm{AR}$ for that event. Horizontal or vertical bars in the shading would imply magnetic models at one particular time could be responsible for the biases seen in Table \ref{tab:gammatable}; while some bars can be seen, they do not appear to underlie the predominance of positive $\Delta \gamma_\textrm{AR}$ values in this Table. The smoothness in the shading for most events demonstrates that typical evolution in $\gamma_\textrm{AR}$ is steady. Looking at all 10 panels, there is a slight predominance of blue, consistent with the bias seen in $\Delta \gamma_\textrm{AR}$ in Table \ref{tab:gammatable}. 

Recall that a positive $\Delta \gamma$ corresponds to a less-steep decrease of magnetic field strength with radius. Decreasing confinement would correspond to $\gamma$ becoming increasingly negative prior to a CME. However, only one event (June 12, 2005) clearly showed this behavior.


\subsection{Evolution in $B_{t}$}
\label{sec:btransverse}

\begin{deluxetable}{lrrrrr}
\tabletypesize{\scriptsize}
\tablecaption{\hspace{-.07in}Evolution of Average Transverse B Field Strength at Height Closest to 42 Mm}
\tablewidth{\columnwidth}
\tablehead{
  \colhead{\hspace{-.07in}Initial Time\tablenotemark{a}}
& \colhead{\hspace{-.1in}Final Time\tablenotemark{a}}
& \colhead{\hspace{-.1in}$\Delta {B_t}_\textrm{AR}>0$\tablenotemark{b}}
& \colhead{\hspace{-.1in}$\Delta {B_t}_\textrm{AR}<0$}
& \colhead{\hspace{-.1in}$\Delta {B_t}_{\textrm{PIL}}>0$}
& \colhead{\hspace{-.1in}$\Delta {B_t}_{\textrm{PIL}}<0$}  
}
\startdata
-6
& 0
& 5
& 5
& 7
& 3\\
-12  
& 0
& 6
& 4
& 6
& 4\\
-24
& 0
& 6 (1)
& 4
& 4 (1)
& 6\\
-48
& 0
& 5 (1,1)
& 5
& 7 (2,1)
& 3\\
-12
& 12
& 6
& 4
& 6
& 4\\
-24
& 24
& 6 (2)
& 4
& 4 (2)
& 6\\
-48
& 24
& 5 (2,1)
& 5
& 7 (3,1)
& 3\\
\enddata 
\tablenotetext{a}{Measured in hours since closest magnetogram to CME event}
\tablenotetext{b}{Most values are not significant at a 1$\sigma$ confidence level; the number of events for which values are significant are listed in parentheses with the first and second entries corresponding to 1$\sigma$ and 2$\sigma$ significance, respectively.}
\label{tab:barraytable}
\normalsize
\end{deluxetable}

\citet{Liu2008} also suggests weaker transverse field strengths at 42 Mm more easily enable CMEs. To test if changes in $B_t$ are related to CME onset, we also looked at evolution in transverse field strength at 46.5 Mm, the height in our models closest to 42 Mm, which Liu used in this aspect of his modeling. Looking at evolution of $B_t$ plotted for each event in Figure \ref{fig:bevolution}, we see no typical trend in the evolution of field strengths surrounding our CMEs. For December 12 2007 we see a slight increase in $B_{t}$, while for May 12, 1997 we see a slight decrease in $B_{t}$. A key point here is that CMEs can evidently occur for any type of evolution in $B_{t}$, as with evolution in $\gamma$. Furthermore, we found no obvious correlation between $B_t(t_0)$ and either CME speed or flare size (see the right three columns of Table \ref{tab:artable}). We saw that PILs had a higher transverse magnetic field compared to those of whole ARs. This is probably due to the structure of the magnetic field at the PIL: tracing field lines between two polarities, the potential magnetic field generally runs nearly parallel to the surface above the PIL, leading to a comparatively high value of transverse field strength. As with $\Delta \gamma$, we computed several $\Delta B_t$'s, as shown in Table \ref{tab:barraytable}. There is a slight predominance of positive $\Delta B_{t\textrm{PIL}}$'s --- five out of seven cases --- which would have a likelihood of 0.07 if $\Delta B_{t\textrm{PIL}}$ were random, and obeyed the binomial distribution. As with $\Delta \gamma_\textrm{AR}$, we calculated changes in normalized $\Delta B_{t\textrm{PIL}}$ for all time intervals, and plotted these differences in Figure \ref{fig:trianglebarraypil} with  the same shading and scaling. There is a slight predominance of blue, corresponding to a tendency for increases in $B_t$. We note that there is also a predominance of positive $\Delta B_{t\textrm{AR}}$'s in Table \ref{tab:barraytable}, but not as great as that among the $\Delta B_{t\textrm{PIL}}$'s.

To see if increases in $B_t$ could be interpreted as due to flux emergence, we plotted the total unsigned radial flux of the entire AR for each model for each event in Figure \ref{fig:unsignedflux}. Error bars were computed assuming 20 Mx cm$^{-2}$ per MDI pixel, or 2.5 $\times 10^{20}$ Mx for each $\sim 12$ Mm PFSS pixel. With some exceptions, there is a strong correlation between the evolution of the average $B_t$ at 46.5 Mm and the total unsigned flux. While increases in $B_{t}$ in many cases appear related to flux emergence (e.g., March 13, 2003, and Nov. 06, 2004), there are also clear counterexamples (e.g., Jan. 03, 2005). Also, we found examples of CMEs occurring during episodes of both increases and decreases in flux (e.g., Dec. 12, 2007 and Jan. 03, 2005, respectively). Observations of CMEs while flux is both increasing and decreasing imply that changes in flux, by themselves, are not necessary conditions for CMEs to occur. We remark that \citet{Mason2010} found flux emergence to be a less powerful indicator of large flares (cf., CMEs) than other properties of photospheric fields. 

We note that uncertainties for this total unsigned flux do not include systematic uncertainties in the model radial magnetic fields, such as assumptions \citet{Schrivjer2003} made in converting the line-of-sight magnetic field to the radial magnetic field, the spatial and temporal sampling of the line-of-sight magnetic field measurements, Fourier ringing from the spherical harmonic expansion, and a well known saturation effect in the MDI instrument that underestimates flux densities in strong field regions \citep{Liu2007}.\\


\subsection{Evolution of Overlying Flux at Larger Radii}
\label{sec:largeradii}

Following \citet{Wang2007b}, we sought to investigate the amount of transverse flux overlying PILs. They computed transverse-field fluxes over low (from 1.0 to 1.1$R_\odot$) and high (from 1.1 to 1.5$R_\odot$) radii crossing planes parallel to their PILs, as well as the low-to-high ratio of these two fluxes. We numerically integrated the transverse field strength over radius in these low and high ranges for all PIL pixels, averaged the integrated quantities over all the PIL pixels, and took the low-high ratio. Each term in the ratio therefore had units of Mx cm$^{-1}$, and can be thought of as the average transverse flux per pixel.

We note that their PFSS model fields incorporated more spherical harmonic coefficients, with $\ell_{\rm max} = 225$, while the default value of $\ell_{\rm max}$ for our models is 192. In principle, over the low radii from 1.0 to 1.1 $R_\odot$, higher-$\ell_{\rm max}$ models can more accurately represent the potential magnetic field from a given boundary condition. It is also true, however, that the corona is thought to be increasingly non-potential at lower heights. Consequently, a higher-order potential-field model does not necessarily more accurately represent the actual field in the low corona. To check if high-low flux ratios depend strongly on $\ell_{\rm max}$, we computed the ratios with two $\ell_{\rm max}$ values, 192 and 225.

\begin{figure*}[!ht]%
\centerline{%
\includegraphics[width=.9\textwidth]{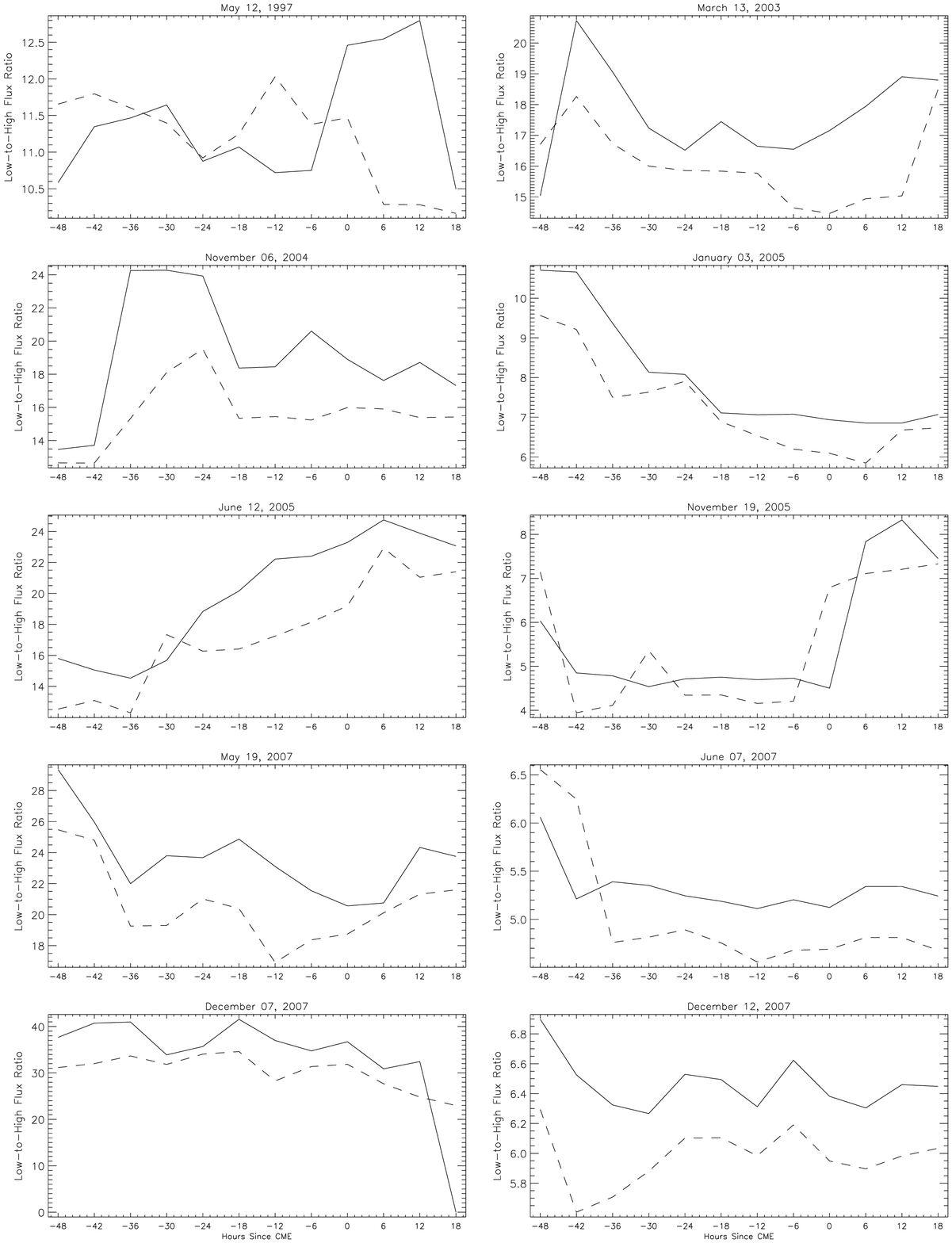}}
\caption{Ratio of radially-integrated transverse field ranging between heights of 1.0--1.1$R_\odot$ to radially-integrated field between 1.1--1.5$R_\odot$. Transverse field strengths with $\ell_{\rm max}=225$ are plotted solid while field strengths with $\ell_{\rm max}=192$ are plotted dashed.}
\label{new_ratio_pil_plots}
\end{figure*}

In Figure \ref{new_ratio_pil_plots}, we plot time series of the low-to-high flux ratio, for two choices of the maximum spherical harmonic coefficient of $\ell_{\rm max}$, 192 (dashed) and 225 (solid). A small low-high ratio implies weaker low-lying transverse fields relative to transverse fields at greater heights. 
Comparing trends between events, no evidence for regular behavior is seen for both values of $\ell_{\rm max}$. For instance, in some events the ratio decreases (e.g., Jan. 03, 2005), while for others it increases (e.g., June 12, 2005). 

The computations with  $\ell_{\rm max} = 225$ were undertaken to check if using higher $\ell$ terms in the spherical harmonic expansion strongly affected the low-high ratios. We found similar evolution in the low-to-high ratio for either value of $\ell_{\rm max}$: for all events except the May 12, 1997 event, the low-high ratios for each $\ell_{\rm max}$ are significantly correlated over our 13 sample times, with rank-order correlation coefficients in the range 0.4 -- 0.9. 

For the May 12, 1997 event, we qualitatively compared the reconstructed photospheric radial fields for both values of $\ell_{\rm max}$ at $t_0$, and found that the higher-$\ell$ radial field had more salt-and-pepper weak fields, which would increase the average transverse field strength at low heights. The discrepant evolution of the low-to-high ratio for this event does not change our conclusion that evolution in the low-to-high flux ratio does not exhibit any consistent trends around CME onset times.

\begin{figure}[!h]%
\vspace{5pt}
\centerline{%
\includegraphics[width=4.5in]{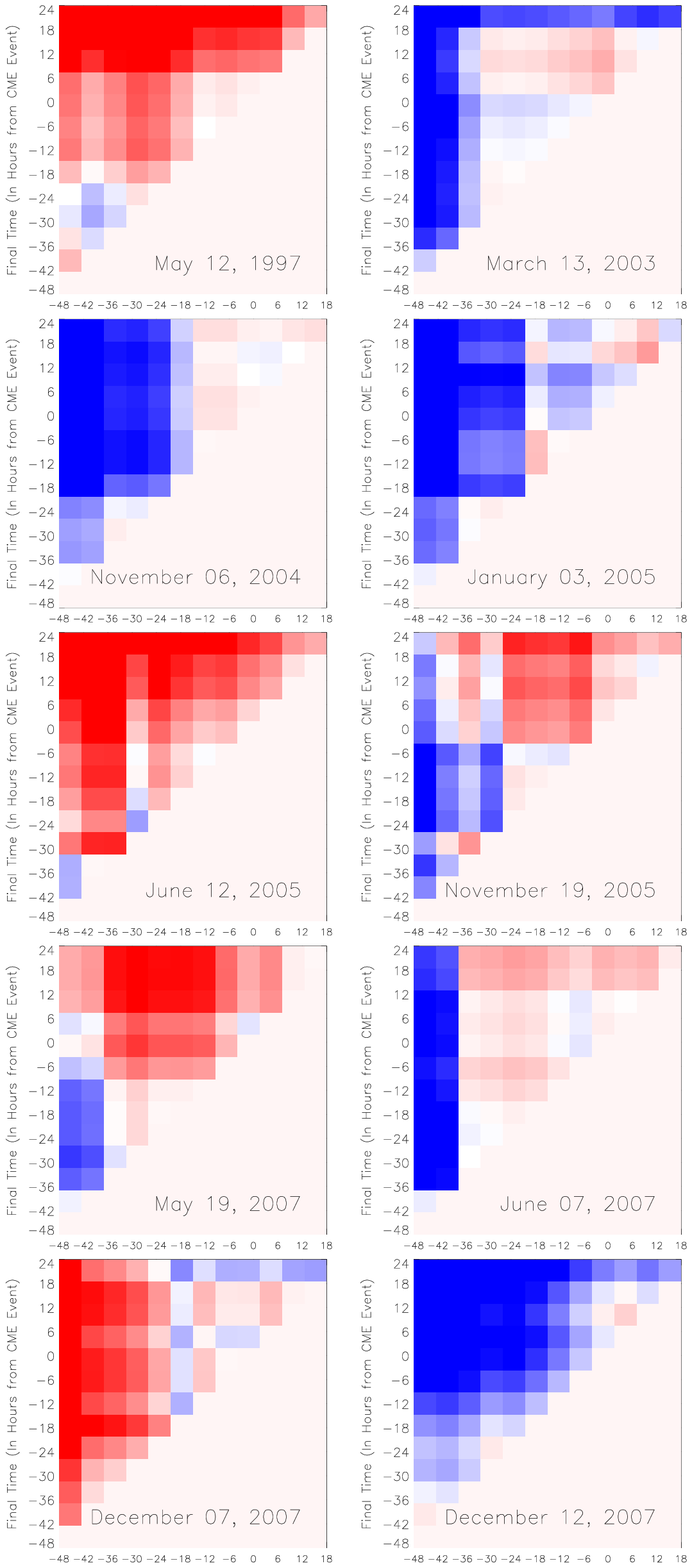}}
\vspace{10pt}
\caption{Differences of overlying integrated fields, $\Delta \Phi$, between all possible pairs of $\Phi(1.1-1.5R_{\odot})$. The saturation level is $\pm 2$, scaled by the standard deviation in $\Phi(1.1-1.5R_{\odot})$ for each event, with negative changes represented as red and positive changes as blue.}
\label{tab:highflux} 
\end{figure}

In addition to studying evolution of the ratio of low-to-high fluxes, we also considered changes in the high flux, by itself. In Figure \ref{tab:highflux} we show pairwise time differences $\Delta \Phi$ of radially-integrated fields from $1.1-1.5R_{\odot},$ which we denote $\Phi(1.1-1.5R_{\odot})$, for all possible time pairs for each event, analogous to differences shown in Figures \ref{tab:trianglegamma} and 7. For most events, these plots behave similarly to those of $B_{t\textrm{PIL}}$ in Figure 7: trends in the magnetic field starting at 46.5 Mm continue at higher altitudes up to 1.5$R_\odot$. For the June 12, 2005 event, however, $B_t$ at 46.5 Mm increases, while the integrated field from 1.1 to 1.5$R_\odot$ decreases. We note that the decreasing integrated field at high radii is consistent with the more-negative $\gamma$ for this event in Figure \ref{fig:finalplotsv2}.


\section{Discussion}
\label{sec:conclusion}

Pre-CME coronal fields are thought to be in a state of balance between outward-directed magnetic pressure and inward-directed magnetic tension. Changes in the large scale magnetic environment of a pre-CME magnetic field could reduce the confining tension force, causing an imbalance, and thereby initiating the CME process. Given rapid time scales of magnetohydrodynamic instabilities in the corona, any significant changes in either outward pressure or inward confinement could lead to rapid CME onset. For instance, \citet{Kliem2006} proposed that if the decay with radius of overlying magnetic field is sufficiently rapid, then the torus instability would occur. 

We looked for evidence of significant evolution in the large-scale magnetic environment in ten CME source regions from about two days before to one day after each event. We studied isolated active regions on the sun and the magnetic field decay rates above these regions as functions of time. We masked subsets of pixels from magnetograms into two categories: active regions (ARs) and polarity inversion lines (PILs). For each of these two categories we applied two separate methods of computing the power-law exponent, $\gamma$, characterizing the decay rate of transverse field with radius above the model photosphere. In Method 1, we averaged transverse magnetic field strength at a given radius over all pixels in the mask, then fitted the logarithm of these averages against the logarithm of radius to find $\gamma$. In Method 2, we fitted a power-law exponent $\gamma$  for the decay of $B_t$ with $r$ for each pixel within a mask (AR or PIL) and then averaged these $\gamma$'s. Even though the values of $\gamma$ were different for methods 1 \& 2, a more important factor that affected the fitted $\gamma$ was which pixels were being used in the mask: AR versus PIL. 

We then sought any relation between evolution of radial decay rates of $B_t$ and CME onset. On no particular time scale is there a statistically significant difference in the number of CME events whose $\gamma$ increased versus decreased. On the contrary, the majority of $\gamma$'s maintained a steady value with variations that failed to exceed a $1-\sigma$ confidence level. That said, while no strong trends were present, we did find a slight tendency for $\gamma$'s to become less negative, consistent with {\em increased} transverse field strengths at higher altitudes around CME times. 

\citet{Liu2008} suggested transverse magnetic field strengths, at heights of approximately 42 Mm, may quantify the strength of magnetic confinement, implying that changes in this field could be related to CME onset. Accordingly, in addition to modeling $\gamma$, we computed the PIL-mask- and AR-mask-averaged transverse field strengths at 46.5 Mm above our CME source regions. As shown in Figure \ref{fig:bevolution}, we also saw no common trends in any time scale for this magnetic parameter.

In addition, we found that CME speeds and strengths of their associated flares were not significantly correlated with either $\gamma$ or the average overlying transverse magnetic field strength at 46.5 Mm.

\citet{Wang2007b} analyzed the ratio of transverse fluxes overlying flaring PILs over low (1.0--1.1$R_\odot$) and high (1.1--1.5$R_\odot$) altitudes. They found regions with lower ratios of low-to-high flux were less likely to erupt. We looked at radially integrated transverse fields over these same ranges and found no evidence for significant, systematic changes in our sample, implying that no particular evolution of the low-to-high flux ratio is common near the times of CMEs.

We find that evolution in the field-strength decay parameter $\gamma$, average transverse field strength $B_t$ near 50 Mm, and radially integrated overlying field do not follow any obvious patterns around CME onset in our potential-field models of source regions. 
Phrased another way, we find that evolution in our parametrizations of large-scale structure in PFSS models is not, by itself, a necessary condition for ejections to occur: eruptions were observed to occur whether our parameters increased, decreased, or stayed the same. If our potential field models accurately represent the large-scale field in CME source regions, then our results imply that no particular large-scale evolution is, by itself, necessary for CMEs to occur.

We expect, however, that large-scale field evolution could play an important role in the CME process, albeit one that varies depending upon evolution in the core fields of CME source regions. For instance, it is possible that the torus instability (or any other instability related to the rate of decay of field strength with radius) might play key roles in the CME process, but our results suggest that crucial evolution occurs at heights lower than those we modeled (primarily 46.5 Mm and above). Both the low spatial resolution of our PFSS models and their neglect of the coronal electric currents that drive CMEs, however, imply that we cannot sensibly extend our analysis to smaller radii. 

We remark that our models cannot accurately capture rapid changes in confinement. For instance, our models do not address the possibility of decrease in confinement via reconnection in the overlying fields, as in the breakout model \citep{Antiochos1999a}. Investigating such processes would require a global-scale, dynamic field model.

\citet{Liu2008} and \citet{Wang2007b} compared potential fields above sites of both eruptive and non-eruptive flares, and found that their parametrizations of large-scale confinement were weaker in eruptive flares. This suggested both that potential models could quantify confinement, and that the strength of confinement is a key factor in the eruption process. We studied the evolution of similar parametrizations of confinement prior to several CMEs, and saw no clear patterns in these parameters' behavior around the time of each CME. Because we did not compare regions with and without ejections, our results are not necessarily inconsistent with previous results. We conclude, however, that the measures of confinement that we have investigated are of limited utility in forecasting CMEs.


\acknowledgements We gratefully acknowledge support from the NSF's National Space Weather program, under award \# AGS-1024862. Magnetogram data is taken from MDI, a project of the Stanford-Lockheed Institute for Space Research and a joint effort of the Solar Oscillations Investigation (SOI). Also, we thank Marc DeRosa for his help with the PFSS Models and re-running the December 2007 CMEs due to an earlier bug error in the model.

\eject

\bibliographystyle{apj}
\bibliography{abbrevs,%
short_abbrevs,%
fuller_lib}

\begin{thebibliography}{20}
\expandafter\ifx\csname natexlab\endcsname\relax\def\natexlab#1{#1}\fi

\bibitem[{Andrews(2003)}]{Andrews2003}
Andrews, M. 2003, Solar Physics, 218, 261

\bibitem[{{Antiochos} {et~al.}(1999){Antiochos}, {DeVore}, \&
  {Klimchuk}}]{Antiochos1999a}
{Antiochos}, S.~K., {DeVore}, C.~R., \& {Klimchuk}, J.~A. 1999, ApJ, 510, 485

\bibitem[{{Falconer} {et~al.}(2006){Falconer}, {Moore}, \&
  {Gary}}]{Falconer2006}
{Falconer}, D.~A., {Moore}, R.~L., \& {Gary}, G.~A. 2006, \apj, 644, 1258

\bibitem[{{Forbes}(2000)}]{Forbes2000}
{Forbes}, T.~G. 2000, JGR, 105, 23153

\bibitem[{Gosling(1993)}]{Gosling1993}
Gosling, J.~T. 1993, JGR, 98, 18,937

\bibitem[{{Kliem} \& {T{\"o}r{\"o}k}(2006)}]{Kliem2006}
{Kliem}, B. \& {T{\"o}r{\"o}k}, T. 2006, Physical Review Letters, 96, 255002

\bibitem[{{Li} {et~al.}(2010){Li}, {Lynch}, {Welsch}, {Stenborg}, {Luhmann},
  {Fisher}, {Liu}, \& {Nightingale}}]{Li2010}
{Li}, Y., {Lynch}, B.~J., {Welsch}, B.~T., {Stenborg}, G.~A., {Luhmann}, J.~G.,
  {Fisher}, G.~H., {Liu}, Y., \& {Nightingale}, R.~W. 2010, \solphys, 264, 149

\bibitem[{{Liu}(2008)}]{Liu2008}
{Liu}, Y. 2008, \apjl, 679, L151

\bibitem[{{Liu} {et~al.}(2007){Liu}, {Norton}, \& {Scherrer}}]{Liu2007}
{Liu}, Y., {Norton}, A.~A., \& {Scherrer}, P.~H. 2007, \solphys, 241, 185

\bibitem[{Longcope {et~al.}(2005)Longcope, McKenzie, Cirtain, \&
  Scott}]{Longcope2005}
Longcope, D.~W., McKenzie, D., Cirtain, J., \& Scott, J. 2005, \apj, 630, 596

\bibitem[{{Martin} \& {McAllister}(1996)}]{Martin1996}
{Martin}, S.~F. \& {McAllister}, A.~H. 1996, in IAU Colloq. 153: Magnetodynamic
  Phenomena in the Solar Atmosphere - Prototypes of Stellar Magnetic Activity,
  497--+

\bibitem[{{Mason} \& {Hoeksema}(2010)}]{Mason2010}
{Mason}, J.~P. \& {Hoeksema}, J.~T. 2010, \apj, 723, 634

\bibitem[{{Moore} {et~al.}(2001){Moore}, {Sterling}, {Hudson}, \&
  {Lemen}}]{Moore2001}
{Moore}, R.~L., {Sterling}, A.~C., {Hudson}, H.~S., \& {Lemen}, J.~R. 2001,
  ApJ, 552, 833

\bibitem[{Press {et~al.}(1992)Press, Teukolsky, Vetterling, \&
  Flannery}]{Press1992}
Press, W.~H., Teukolsky, S.~A., Vetterling, W.~T., \& Flannery, B.~P. 1992,
  Numerical Recipes in C: The art of scientific computing, Second Edition
  (Cambridge: Cambridge University Press)

\bibitem[{Scherrer {et~al.}(1995)Scherrer, Bogart, Bush, Hoeksema, Kosovichev,
  Schou, Rosenberg, Springer, Tarbell, Title, Wolfson, Zayer, \& {The MDI
  Engineering Team}}]{Scherrer1995}
Scherrer, P., Bogart, R.~S., Bush, R.~I., Hoeksema, J.~T., Kosovichev, A.,
  Schou, J., Rosenberg, W., Springer, L., Tarbell, T., Title, A., Wolfson, C.,
  Zayer, I., \& {The MDI Engineering Team}. 1995, Solar~Phys., 162, 129

\bibitem[{{Schrijver}(2007)}]{Schrijver2007}
{Schrijver}, C.~J. 2007, \apjl, 655, L117

\bibitem[{{Schrijver} \& {De Rosa}(2003)}]{Schrivjer2003}
{Schrijver}, C.~J. \& {De Rosa}, M.~L. 2003, \solphys, 212, 165

\bibitem[{{Schrijver} {et~al.}(2005){Schrijver}, {DeRosa}, {Title}, \&
  {Metcalf}}]{Schrijver2005}
{Schrijver}, C.~J., {DeRosa}, M.~L., {Title}, A.~M., \& {Metcalf}, T.~R. 2005,
  \apj, 628, 501

\bibitem[{{Wang} \& {Zhang}(2007)}]{Wang2007b}
{Wang}, Y. \& {Zhang}, J. 2007, \apj, 665, 1428

\bibitem[{{Welsch} \& {Li}(2008)}]{Welsch2008b}
{Welsch}, B.~T. \& {Li}, Y. 2008, in Astronomical Society of the Pacific
  Conference Series, Vol. 383, Astronomical Society of the Pacific Conference
  Series, ed. R.~{Howe}, R.~W. {Komm}, K.~S. {Balasubramaniam}, \& G.~J.~D.
  {Petrie}, 429--437; also arXiv:0710.0562

\end{thebibliography}

\end{document}